\documentclass[numreferences,oneside,final]{kluwer}
\usepackage{graphicx}

\newcommand\mean[1]{{\left\langle#1\right\rangle}}
\begin{document}

\begin{article}
\begin{opening}

\title{Ordering at Solid-Liquid Interfaces Between Dissimilar Materials}

\author{Adham \surname{Hashibon}, }
\author{Joan \surname{Adler}, }
\institute{Department of Physics, Technion -- Israel Institute of Technology, Haifa, Israel}

\author{Michael W. \surname{Finnis}, }
\institute{Atomistic Simulation Group, School of Mathematics and Physics.
 The Queen's University of Belfast, Belfast BT7 1NN}

\author{Wayne D. \surname{Kaplan}.}
\institute{Department of Materials Engineering, Technion, Haifa, Israel.}

\runningauthor{Adham Hashibon}
\runningtitle{Ordering at Solid-Liquid Interfaces Between Dissimilar Materials}

\date{July 23, 2001}

\begin{abstract}
In an earlier report we explored structural correlations at a liquid-solid
interface  with molecular dynamics simulations of a model aluminium system
using the Ercolessi-Adams potential and up to 4320 atoms. Substrate atoms were
pinned to their equilibrium fcc crystalline positions while liquid atoms were
free to move. A direct correlation between the amount of ordering in the liquid
phase and the underlying substrate orientation was found. In the present paper
we extend this study to  the case of a fixed bcc substrate in contact with
liquid aluminium. We find surprisingly similar   results for the density profiles
of both (100) and (110) substrates. However, there is a far greater in-plane
ordering in the (100) than for the (110) system. For the (100) substrates we
observe adsorption of liquid atoms into the terminating plane of the bcc (100)
substrate, effectively transforming the bcc (100) plane into an fcc (100) plane.
\end{abstract}


\end{opening}
\centerline{\sf Accepted by Interface Science, September 2001}

\section{Introduction}
\label{Introduction}
Macroscopic properties of metal-ceramic interfaces are strongly
correlated with microscopic details of the metal-ceramic interface
such as wetting, chemistry, diffusion, and structure.  Correlating
macroscopic properties to the structure and chemistry of interfaces is
one of the most intriguing topics in materials science.  Experimental
studies of the atomistic structure of a solid-{\it liquid} internal
interface are technically difficult to
conduct~\cite{Huisman1997,Reichert2000}, and {\em ab-initio}
simulations are limited in size. Therefore, atomistic simulations of
metal ceramic interfaces can serve as an important tool to understand
and predict the effect of the interface region on the material
properties.

Atomistic simulations, such as Molecular Dynamics (MD) or Monte Carlo
permit the controlled study of these systems at the atomistic level
for a large number of atoms and for large structures. However the main
limitation to such simulations is the lack of appropriate interatomic
potential schemes which can model both metallic and ionic bonding
across the interface. Nevertheless, simplified models can be used to
obtain qualitative basic insights into the problem.  In an earlier
study~\cite{paper1} we introduced a model system in which the ceramic
is assumed to be composed of atoms pinned to their equilibrium lattice
positions, while the metal atoms are free to evolve under the
influence of their interatomic potential.

The atoms of a liquid metal which are adjacent to a rigid crystalline
substrate are in an environment which is strongly affected by the
symmetry of the underlying substrate. Theoretical
studies~\cite{Howe1996}, which are mainly computational, have shown
that ordering occurs in the first layers of the liquid adjacent to the
crystal surface.  The same result emerges from experimental studies of
solid-liquid interfaces~\cite{Huisman1997,Howe1996}. The layering is
seen to decay exponentially with increasing distance from the
substrate. In \cite{paper1} simulations were conducted to study the
density profile and structure of the liquid-metal/hard-wall interface
as a function of temperature and substrate structure for an fcc
substrate with aluminium liquid. We found that the decay of the
density profile is quantitatively and qualitatively related to the
underlying structure of the substrate.

In this paper, we address the issues that arise when the substrate is switched
to a bcc structure. Since a bcc substrate and liquid aluminium have rather
incompatible structures, this should have more general implications.

\section{System and Simulation Method}
\label{sec:system}
We use the same simulation method as \cite{paper1}, with several layers
of fixed substrate in the ideal lattice positions, and the remaining
atoms free to move. The potential~\cite{EA1994} is the same aluminium
potential used and verified in~\cite{paper1}, and all computational aspects are
similar.

In the plane of the interface (the $xy$ plane) periodic boundary
conditions were applied.  In the direction perpendicular to the
interface (the $z$ direction), the boundary conditions are expected to
simulate the bulk media on either side of the interface.  On the rigid
(substrate) side where atoms are fixed to crystalline positions, it is
sufficient to require that the extent of the region in the $z$
direction to be larger than the cutoff of the interaction potential.
In this way liquid atoms near the interface do not ``see'' the bottom
of the rigid layer, hence it acts like a semi infinite bulk system.
The fcc substrate had the lattice parameter of $4.1$\AA, close to that
of solid Al, and the bcc substrate was given the same lattice
parameter so that the bcc substrate was less dense by a factor of
half.

A liquid layer is deposited above the solid layer, and then a vacuum
region is inserted with periodic boundary conditions in all
directions. In this way we have one free liquid interface and one
internal solid-liquid interface. Provided that the height of the
liquid layer is large enough in the $z$ direction, there will be no
interaction between the free liquid surface and the internal
rigid-liquid interface. In addition the system will be free to respond
to stresses at the interface since there is nothing to limit the
liquid from expanding in the $z$ direction.

The system was simulated using an MD technique, which consists of the
numerical integration of Newton's equation of motion for the 
atoms~\cite{FS1996}. The velocity-Verlet integration
algorithm~\cite{FS1996,AT1987} was used in the simulations.  For the
interatomic potential, an embedded atom potential developed by
Ercolessi and Adams (EA)~\cite{glue} was used. This potential was
constructed by the so called force matching method, whereby the
potential was fitted to a very large amount of data obtained from both
experiment and first principle calculations, with emphasis given to
match the interatomic forces obtained from the potential to those
obtained from first principles. The potential has been tested in
detail for aluminium\cite{EA1994,Hansen99}, and was found to be
consistent with experimental results. For example, the calculated
melting point for aluminium is T=939$\pm$5K in excellent agreement
with the experimental value of T=933.6K.

The density profile $\rho(z)$ is defined as the average density of particles in a slice of width $\Delta z$ parallel to the hard wall surface and centered around $x$. The simulation cell is divided into equal layers or bins parallel to the interface.  The expression for the density profile is
\begin{equation}
\label{eq:rho}
\rho(z) = \frac{\mean{N_z}}{L_xL_y\Delta z} \nonumber
\label{eq:N_x}
\end{equation}
where $L_x$ and $L_y$ are the $x$ and $y$ dimensions of the cell,
respectively, and $z$ is perpendicular to the interface, $\Delta z$ is
the bin width, and $N_z$ is the number of particles between $z-\Delta
z/2$ and $z+\Delta z /2 $ at time $t$. The angled brackets indicate a
time average.

In order to reduce the statistical error of the sampling, a proper
choice of bin width must be made. Very small bin widths results in too
few particles at each time step, hence a large scatter of the data.
Very large bins will not show the actual dependence of the density
profile over distance.  Two basic width scales have been used: a
coarse scale, in which the width of the bins was set equal to the bulk
crystal d-spacing (the distance between consecutive planes) for a
particular orientation, and a fine scale for which each coarse scale bin was
divided into 10 or 25 sections.

The upper envelope of the density profile is fitted assuming an
exponential decay: $\rho(z) \sim \exp{(-\kappa z)}$, where $\kappa =
1/\xi$, and $\xi$ is the correlation length at the interface.  The
parameter $\kappa$ can then be used to {\em quantitatively} describe
the amount of {\em disorder} at the interface. The exact function that
has been used in the fitting has an extra constant term, $b$, to
account for the background liquid density, and a normalization factor,
$a$:
 \begin{equation}
 \rho(z) = a\rm{e}^{-\kappa z} + b
\label{eq:denprof}
\end{equation}
This form of decay of the density profile is typically obtained from a mean
field treatment of binary fluid interfaces~\cite{Tomagnini96,tarazona}, and was
also validated computationally in~\cite{paper1}.

\section{Results and Discussion}
\label{sec:results}
\begin{figure}[htbp]
  \begin{center}
   \centering\includegraphics[width=8.0cm]{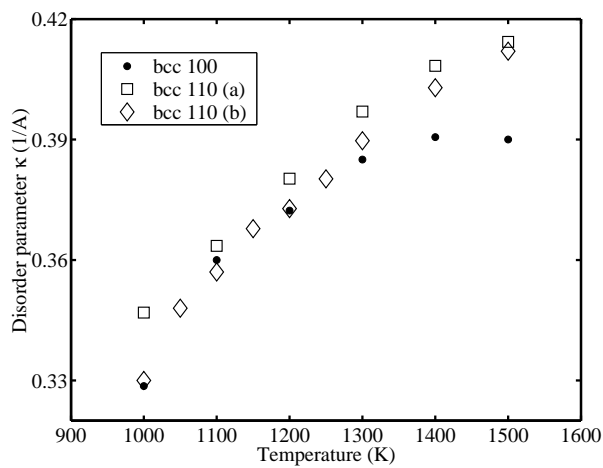}
    \caption{\small The disorder parameter $\kappa$ for bcc substrates. The
      results are very similar for both the (110) and (100) substrates
      at $T\leq 1300K$.}
\label{fig1}   
\end{center}
\end{figure}

Ordering of a liquid near an interface with a solid has two
components; one is related to the extent of the layering, and the
other to the disposition of the atoms within each layer. The former
can be described quantitatively by extracting the disorder parameter
$\kappa$ from the decay of the density profile. Figure~\ref{fig1}
shows the disorder parameter for the samples used in this study (the
error bars in the data are of the size of the symbols used). The data
for the (100) system is obtained from a sample composed of 2250 atoms,
with a substrate containing 10 (100) bcc planes. Two sets of data
points for the bcc (110) system are displayed, each is from a
different system; with system {\bf (a)} being smaller than system {\bf
  (b)}. System {\bf (a)} is composed of 1300 atoms with a substrate
made of 6 (110) planes. System {\bf (b)} is composed of 4508 atoms,
also with a substrate made of 6 (110) planes.  For the smaller system
$\kappa$ is consistently {\em larger}, indicating that finite size
effects (FSE) exist in the system. A full finite size scaling analysis
is required to accurately detemine the values of $\kappa$, however
such an analysis is beyond the scope of this work. The shift
due to the FSE is considered to be an additional source of error on the
data, in other words, only when the difference between two values of
$\kappa$ exceeds the shift due to the FSE, then those two values are
considered to be different.

For low temperatures (approximately $T< 1300$K) there is no
substantial difference in $\kappa$ between all systems, i.e. the
amount of disorder exhibited at each interface is the same. However,
for higher temperatures ($T>1300$K) the (100) and (110) systems
deviate substantially, and the (100) system is seen to preserve more order
at higher temperatures than the (110) system.

\begin{figure}[htbp]
  \begin{center}
    \includegraphics[width=8.0cm]{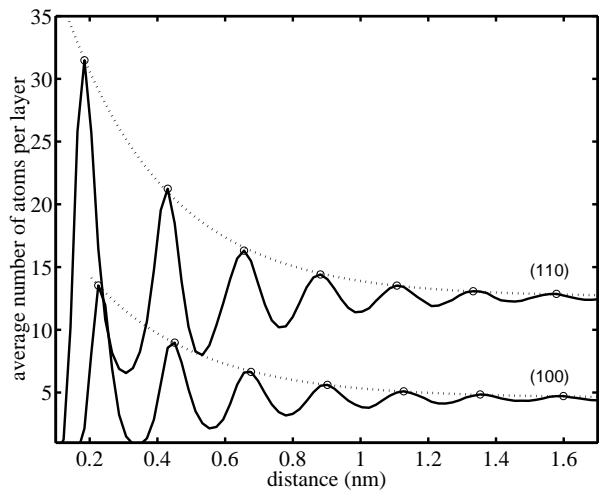}
    \caption{\small  Layering in the liquid  for bcc (100) and bcc (110) substrates at 1000K. The graph  shows the average number of atoms per layer as a function of distance (in the z direction) in nm from the substrate. Only the
      part of the liquid adjacent to the solid is shown, and the
      average number of atoms in each layer is plotted as a function
      of the distance from the solid substrate.}
\label{fig2} \end{center}
\end{figure}
The proximity of $\kappa$ at low temperatures can be further
demonstrated from the comparison of the density profiles of the two
systems in Figure~\ref{fig2}. The average number of atoms in each
layer is plotted as a function of the distance of the layer from the
solid substrate at a temperature of $T=1000$K. The peaks corresponding
to the solid phase are not shown.  Note that the y-axis is the
average number of atoms found in each layer and not the {\em density}.
The layer magnitude, or average number of atoms in each layer, decays
gradually to a constant value. When this value is normalized by the
volume of each layer, it will be equal to the density of the liquid,
which is the same for all systems. The dashed lines are an actual
fit to Equation~\ref{eq:denprof}. Despite  the difference in the
magnitude and the distance from the substrate, both profiles are very
similar, with the same number of density peaks, and a very similar
envelope (dashed lines). Hence it is not surprising that $\kappa$ is
very close for both systems.

Since both the in-plane structures and the d-spacings of the bcc (110)
and bcc (100) substrates are different, we would expect that some of
this difference be reflected in the structure of the liquid at the
interface.  However we obtain very similar results for both $\kappa$,
and for the density profiles.

\begin{figure}[htbp]
  \begin{center}
    \includegraphics[width=8.0cm]{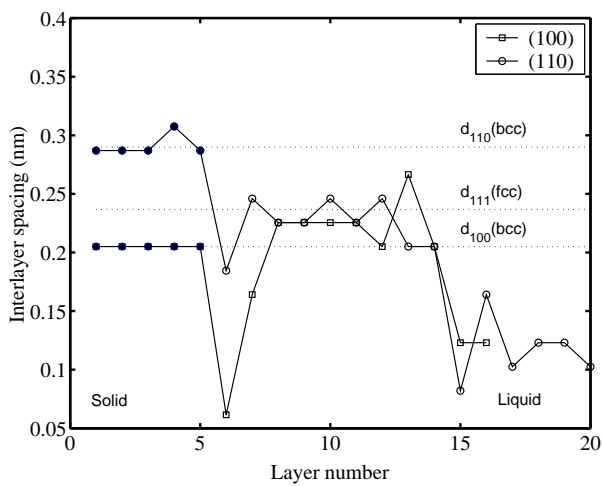}
    \caption{\small The d-spacing for bcc (100) and bcc (110) substrates;
    in both cases the interlayer spacing inside the liquid is compatible
    with the periodicity of  (111) planes in an fcc metal.}
    \label{fig3}
  \end{center}
\end{figure}

We next examine the interlayer spacing for both systems, as shown in
Figure~\ref{fig3}. The closed symbols indicate the d-spacing
found in the ideal structure (the small solid peak in the case of the
(110) system is due to a small shift in the slicing). In both systems
the interlayer distance of the quasi-liquid is {\em uncorrelated} with
the d-spacing of the substrate. This suggests that in both cases the
in-plane structure in the liquid is not similar to that of the
substrate. Moreover, apart from first layer in each system, the
interlayer spacing in both cases is seen to fluctuate around a value
which is close to the d-spacing of a (111) family of planes in an
ideal {\em fcc} material.  The drop in the d-spacing of the first
layer for the (110) system is mainly an artifact of the slicing scheme
as noted earlier. The large drop in the d-spacing of the first layer
for the (100) system is however, related to a physical phenomenon that
is explained below.

\begin{figure}[htbp]
  \begin{center}
    \includegraphics[width=8.0cm]{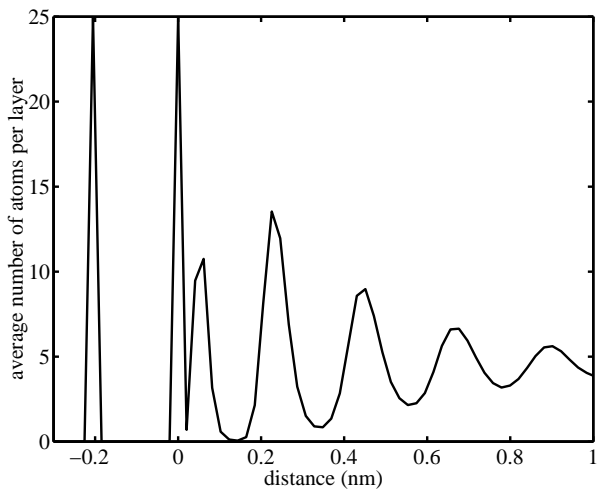}
    \caption{\small The profile of the bcc (100) system showing an extra peak very near to the solid peak, which is due to the adsorbed liquid layer.} 
    \label{fig4}
  \end{center}
\end{figure}

\begin{figure}[htbp]
  \begin{center}
    \includegraphics[width=8.0cm]{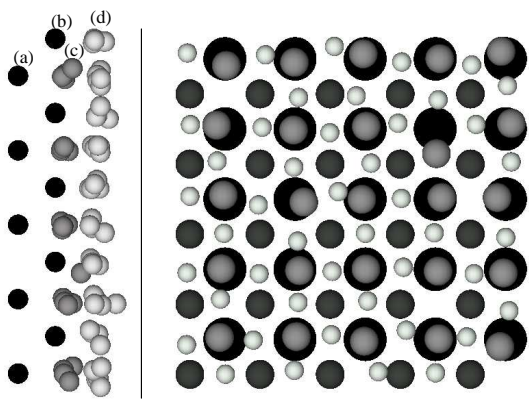}
    \caption{ \small The interface region at a bcc (100) substrate at  $T=1000K$.
      The left panel is a side view and the right panel is a top view.
      The darkest atoms represent the solid, medium and light gray,
      the liquid. On the right, different sizes indicate layer
      position: layer (a) is represented by large spheres, layers (b)
      and (c) by normal size, and layer (c) by a smaller size.  Liquid
      atoms (dark gray) in layer (c) are adsorbed into the solid, they
      lie in the {\em face-centered} sites on the bcc (100) plane, but
      positioned slightly above the (100) plane as seen from the left
      panel.}
    \label{fig5}
   \end{center}
\end{figure}

Figure~\ref{fig4} shows the density layering in the (100)
system, with a closer look at what happens near the rigid substrate.
The first two peaks shown correspond to the solid substrate. In
addition to the layers well inside the liquid, there is an additional
peak in the density profile, very close to last peak of the substrate.
This peak leads to the substantially low inter-layer spacing for the
first peak in Figure~\ref{fig3}. A better understanding of the
source of this peak is achieved by looking at the actual structure of
the interface, as shown in Figure~\ref{fig5}.  The left panel is
a side view, looking down from the [010] direction, of a part of the
interface showing the last two solid layers, labeled {\bf (a)}, and
{\bf (b)}, and the two liquid layers adjacent to the substrate,
labeled {\bf (c)} and {\bf (d)}.  The right panel is a top view,
looking down from the [100] direction.  The size of the atoms drawn in
the right view was changed to make it easier to discern the different
layers. Layer {\bf (a)} is drawn with larger black spheres, followed
by smaller dark spheres for layer {\bf (b)}. It is clear that the
extra peak in Figure~\ref{fig4} is due to the layer {\bf
  (c)} (dark gray atoms in Figure~\ref{fig5}) formed by the
liquid atoms that penetrate - or adsorb - into the terminating plane
of the rigid substrate. These atoms fall into the face-centered
positions of the bcc (100) plane, immediately above the positions of
the atoms in layer {\bf (a)}, in accordance with the fcc interaction
potential of the liquid. In this way, the terminating plane of the bcc
(100) material transforms, via the adsorption of the liquid, into an
{\bf fcc} (100) plane. Note that half of the atoms in this plane are
rigid and the other half slightly displaced above the ideal positions
of the face-center atoms. Hence we expect that the structure of the
liquid at the bcc (100) substrate will be very similar to that found
at an actual fcc (100) substrate.

\begin{figure}[htbp]
  \begin{center}
    \includegraphics[height=8.0cm]{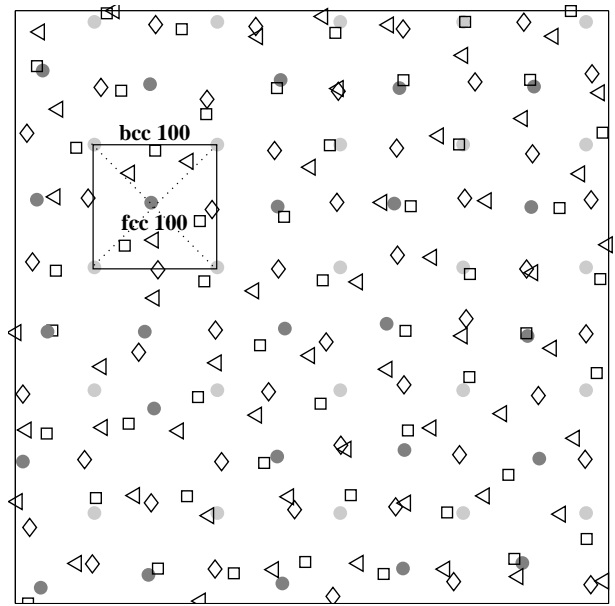}
    \caption{\small Stacking of layers in the bcc (100) system at $T=1000K$.
      Light gray circles represent atoms in the terminating bcc (100)
      plane, dark gray circles the adsorbed layer, diamonds,
      squares, and triangles represent the first three layers in the
      liquid. The square connects four atoms of the bcc (100) plane,
      and the dark circle in the middle indicates the additional site
      of the fcc (100) plane, occupied by an adsorbed atom.}
     \label{fig6}
  \end{center}
\end{figure}
\begin{figure}[htbp]
  \begin{center}
    \includegraphics[width=8.0cm]{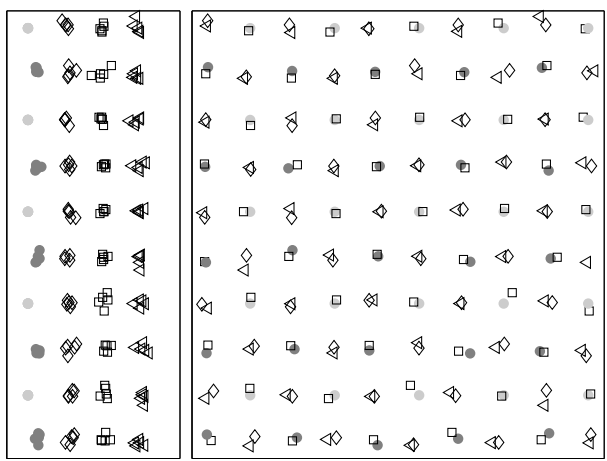}
    \caption{\small Stacking of layers in the bcc (100) system at $T=600K$.
Symbols are the same as for Figure~\ref{fig6}, but the metallic region
has now solidified, forming a bcc (100)/fcc (100) interface.}.
    \label{fig7}
  \end{center}
\end{figure}
In the discussion so far, and in what follows, we assume that the
adsorbed liquid layer is part of the terminating plane of the solid,
so that when calculating $\kappa$ we have excluded this peak from the
density envelope (see Figure~\ref{fig2}).

We now turn our attention to the first liquid layer, labelled {\bf
  (d)} and shown in light gray in Figure~\ref{fig5}. The atoms
of this layer fall into sites directly above holes in the terminating
substrate. There exists a large amount of order in this layer, as can
easily be seen from the right panel of Figure~\ref{fig5}. The
location of the second and third liquid layers is shown in
Figure~\ref{fig6}.  The atoms in the second layer fall into
well defined sites, above the holes in the first layer, and above
atoms in layer {\bf (b)}. In order to see more clearly the actual
stacking of the planes in the liquid phase, we cool down the sample
and let the atoms settle into their equilibrium structures.
Figure~\ref{fig7} shows a snapshot of a sample
solidified at $T=600$K. The stacking of the planes shown and the
position of the atoms indicates the formation of (100) fcc planes.

\begin{figure}[htbp]
\begin{center}
    \includegraphics[width=8.0cm]{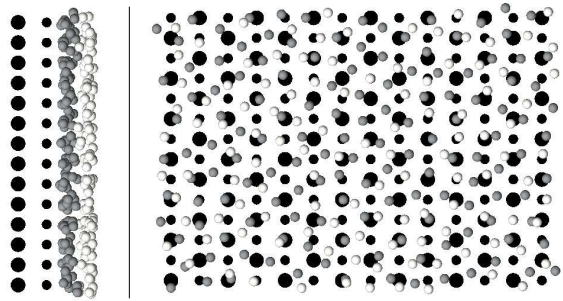}
    \caption{ \small 
      The interface region at a bcc (110) substrate at $T=1000K$.  The
      left panel is a side view and the right panel is a top view.
      The darkest atoms represent the solid, medium and light gray,
      the liquid.  Different sizes indicate layer position as can be
      seen from the left panel.  In contrast to the (100) system,
      there is no adsorption of the liquid to the solid substrate.
      Partial ordering exists in the liquid layers in the manner of a
      (110) system, however many defects are present in the system
      (see text).}
 \label{fig8}
   \end{center}
\end{figure}

On the other hand, the in-plane ordering in the BCC (110) system is
much less pronounced as can be seen from Figure~\ref{fig8},
which shows the structure near the interface at 1000K, the last two
solid planes (darkest atoms), and the first two liquid layers (medium
and light gray). The left panel is a side view, looking down from a
$(0\bar{1}1)$ direction, and the right panel is a top view, looking
down from the (110) direction. It is hard to quantify the amount of
order inside the layers from just looking at the configuration,
although one can clearly see that some order exists, at least short
range order, which resembles a bcc (110) plane. The ordering inside
the planes does not improve much by cooling down the sample.

To summarize, comparing the ordering for both systems, there seems to
be a qualitative difference between ordering normal to the interface,
and ordering inside the planes. Normal to the plane of the interface
the order in both systems is identical, both in terms of $\kappa$ and
of the d-spacing. However, the in-plane structure is strongly
correlated with the underlying substrate, which is very different in
both cases.  In addition, the bcc (100) substrate is strongly affected
by the adsorption of additional aluminium atoms from the liquid phase,
while this effect does not occur with the bcc (110) substrate.
Effectively, we are actually comparing an interface with a fcc (100)
substrate and an interface with a bcc (110) substrate.

Systems with an fcc substrate were investigated in a previous
report~\cite{paper1}, where fcc substrates with three different
terminating planes were simulated: (111), (100), and (110). The
disorder parameter $\kappa$ for the (111) and (100) fcc substrates
varied faster with temperature than the one obtained here for the bcc
substrates. At low temperature, $\kappa$ for the fcc substrates was
smaller, and at high temperature larger, than $\kappa$ for the bcc
substrates. This difference is particularly surprising in view of the
similarity of the fcc (100) and bcc (100) interfaces, which results
from the absorbed layer. At present we have no simple explanation of
the difference.  The disorder parameter, $\kappa$ for the fcc
(110) substrate was much larger than for the other fcc and bcc
substrates, and increased almost linearly with temperature. The
interlayer spacing in the liquid phase for the fcc (111) and (100) was
 very similar to the d-spacing in the bcc systems, that is in all
cases it was very close to the d-spacing of an fcc solid ($d_{111} =
2.367$\AA). The interlayer spacing in the fcc (110) system was equal
to the d-spacing of the underlying substrate. In conclusion,
for the fcc system, the interlayer spacing in the liquid is comparable
to the interlayer spacing inside the fcc substrate.

For the fcc systems, it was also observed that the density changes
gradually to that of the bulk liquid over the interface region. Inside
the interface region there exists a strict correlation between the
interlayer spacing and the in-plane density in the form of the in-plane
density $\zeta$ normalized by the interlayer spacing $d$: 
\begin{equation}
\rho = \zeta/d
\label{eq2}
\end{equation}
where $\rho$ is the bulk density in each layer. For a
perfect lattice this relation is exact, i.e. $\rho$ is the bulk
density of the whole sample. At the interface region $\rho$ 
gradually decays across the interface to the density in the bulk liquid.

For a system with an fcc substrate the difference in density accross
the interface is not large; the density of the fcc substrate being
$\rho_{fcc}=0.058$ atoms/\AA$^3$ (with the lattice paramter of
$4.1$\AA), while that of liquid aluminium within our temperature range
is about $\rho_{Al} = 0.052$ atoms/\AA$^{3}$. Hence there is no
substantial change in the density of the liquid layers, and
consequently there was no major change in the interlayer spacing (see
Equation~\ref{eq2}).

In contrast to the fcc case, the bcc systems are much less dense than
both solid and liquid aluminium (the density of the bcc substrates is
$\rho_{bcc} = 0.029$ atoms/\AA$^3$). The liquid near the interface must
then accomodate for this change in the density across the interface.

When the liquid metal is adjacent to the solid, there will be an
interplay, or competition between the solid and the liquid; the liquid
wants to equilibrate to it's natural density, while the bcc substrates
tend to impose their symmetry on the liquid. In the case of the bcc
(100) substrates, the liquid dominates. The terminating plane of the
bcc (100) is transformed by the ``adsorption of the liquid'' into an
fcc (100) plane, which is much denser than bcc (100). Indeed this
interface behaves similar to one with a true fcc (100)
substrate~\cite{paper1}.

In the case of the bcc (110) system the situation is more complicated.
Again, the liquid tends to order, but the bcc (110) imposes its
in-plane structure onto the liquid.  The system will reach a certain
equilibrium between these two effects. This is exactly what happens in
Figure~\ref{fig8}, where part of the plane is ordered in a (110)
symmetry, and the other is not, while the density in the liquid layers
is typical for liquid aluminium. Table~\ref{tab:tab1} lists the
average in-plane density and the interlayer separation for the bcc
(110) system at a temperature of $T=1000$K. The first layer is from
the perfect lattice, and the rest is the liquid interlayer density.
The values of the perfect in-plane densities for the various surfaces
considered here are also listed for comparison. We note that the
density of the first layer adjacent to the bcc (110) plane jumps to a
value very close to the density of an ideal fcc (111) or (100) plane.
When this density is normalized by the interlayer separation, it is
very close to the bulk liquid density.  This causes the change in the
d-spacing for the (110) system from the bcc (110) value of $d_{110} =
2.8991$\AA, to a value close to an fcc (111).

\begin{table}
\begin{tabular}[h]{ccc|| cc}
\multicolumn{3}{c}{bcc (110)} & \multicolumn{2}{c}{ideal lattices} \\
layer & $d$ & $\zeta$ atoms/\AA$^2$ & plane & $\zeta$ atoms/\AA$^2$ \\
\hline
1.  &  2.8700   &  0.0841  & bcc (100)  &  0.0594 \\
3.  &  2.4600   &  0.1219  & bcc (110)  &  0.0841 \\
4.  &  2.2550   &  0.1202  & fcc (111)  &  0.1374 \\
5.  &  2.2550   &  0.1150  & fcc (100)  &  0.1190 \\
6.  &  2.2550   &  0.1056  & fcc (110)  &  0.0841 \\
    \hline
 \end{tabular}
 \caption{
The width and density of the liquid layers at the interface
with a bcc (110) substrate (left), compared to the planar 
densities for different ideal planes (right).
All models are with a lattice parameter of $a =4.1$\AA.}  
\label{tab:tab1}
\end{table}

\section{Summary and Conclusions}
We have simulated liquid Al in contact with bcc substrates, and
compared the results with similar systems having fcc substrates. We
study the oscillations in the density perpendicular to the interface
and their decay profile as a function of temperature.  The envelope of
the density fluctuation is well described by an exponential,
$a\exp{(-\kappa z)}+b$, where $\kappa$ is a measure of the disorder at
the liquid.  We find that the ordering normal to the plane of the
interface is similar for both systems, despite a very different
structure of the underlying substrates, and that it is dominated by
the properties of the liquid material.  On the other hand, the effect
of the in-plane structure of the substrate is rather pronounced. For
the (100) system, the liquid is much more ordered than for the (110)
system. This is because additional atoms from the liquid phase adsorb
into face centered sites in the bcc (100) surface, forming a structure
resembling solid fcc (100). Work is in progress to understand the
relatively small temperature dependence of the disorder parameter
$\kappa$ on bcc compared to fcc substrates.

{\bf Acknowledgements} A.H. was supported by the Israeli Ministry of
Science.  This research was partially supported by the German-Israel
Science Foundation under grant I-653-181.14/1999, and by the
Binational Science Foundation (Israel-USA, BSF Grants 1998102 and
1999200).

\end{article}
\end{document}